\pdfoutput=1
\documentclass[twocolumn,aps, prl, superscriptaddress]{revtex4-1}
\usepackage{graphicx, amsmath, amssymb, mathtools}
\usepackage[colorlinks=true, citecolor=blue, urlcolor=blue, linkcolor=blue]{hyperref}
\renewcommand{\section}[1]{{\par\it #1.---}\ignorespaces}

\begin{document}
\title{Floquet second-order topological insulators in non-Hermitian systems}
\author{Hong Wu}
\affiliation{School of Physical Science and Technology, Lanzhou University, Lanzhou 730000, China}
\author{Bao-Qin Wang}
\affiliation{School of Physical Science and Technology, Lanzhou University, Lanzhou 730000, China}
\author{Jun-Hong An}
\email{anjhong@lzu.edu.cn}
\affiliation{School of Physical Science and Technology, Lanzhou University, Lanzhou 730000, China}
\begin{abstract}
Second-order topological insulator (SOTI) is featured with the presence of $(d-2)$-dimensional boundary states in $d$-dimension systems. The non-Hermiticity induced breakdown of bulk-boundary correspondence (BBC) and the periodic driving on systems generally obscure the description of non-Hermitian SOTI. To prompt the applications of SOTIs, we explore the role of periodic driving in controllably creating exotic non-Hermitian SOTIs both for 2D and 3D systems. A scheme to retrieve the BBC and a complete description to SOTIs via the bulk topology of such nonequilibrium systems are proposed. It is found that rich exotic non-Hermitian SOTIs with a widely tunable number of 2D corner states and 3D hinge states and a coexistence of the first- and second-order topological insulators are induced by the periodic driving. Enriching the family of topological phases, our result may inspire the exploration to apply SOTIs via tuning the number of corner/hinge states by the periodic driving.
\end{abstract}
\maketitle

\section{Introduction}
Topological insulators (TIs) are intriguing phases that are insulators in their $d$-dimensional bulk but conductors on their $(d-1)$-dimensional boundaries due to the formation of gapless boundary states among the gapped bulk energy bands \cite{RevModPhys.82.3045}. The boundary states are characterized by the topology of bulk bands \cite{RevModPhys.88.035005}, which is called bulk-boundary correspondence (BBC). Recently, TIs have been extended to $n$th-order cases \cite{Benalcazar61,PhysRevB.96.245115,PhysRevLett.119.246401,PhysRevLett.119.246402,Schindlereaat0346,Stefan2018}, which is featured with the presence of $(d-n)$-dimensional boundary states \cite{PhysRevB.97.241405,PhysRevB.99.041301,PhysRevLett.120.026801,PhysRevB.97.205135,PhysRevB.100.235302,PhysRevB.92.085126}. The second-order TI (SOTI) is identified by corner states in 2D systems \cite{PhysRevLett.122.204301,PhysRevLett.122.233902,PhysRevLett.122.233903,PhysRevB.100.205109,PhysRevB.99.165129,PhysRevLett.124.166804,PhysRevLett.123.216803,PhysRevB.101.241104,PhysRevB.101.041109,PhysRevLett.125.036801,PhysRevLett.123.256402} and hinge states in 3D ones \cite{PhysRevLett.121.196801,PhysRevLett.123.247401,PhysRevLett.124.036401,PhysRevB.101.245110,PhysRevB.98.081110,PhysRevB.97.205136,PhysRevLett.123.036802}. With the advances of non-Hermitian physics in optical \cite{PhysRevLett.115.040402,EP2017} and acoustic \cite{PhysRevB.101.180303,PhysRevLett.121.124501} systems, non-Hermitian topological phases have attracted much attention \cite{PhysRevLett.121.136802,PhysRevLett.123.066404,PhysRevLett.123.016805,PhysRevLett.124.056802,PhysRevLett.124.086801,Helbig2020,Weidemann311} due to their diverse applications in laser \cite{Hararieaar4003}, invisible media \cite{Regensburger2012}, and sensing \cite{EP2017,Chen2017}. A unique character of non-Hermitian SOTIs \cite{PhysRevLett.122.195501,PhysRevLett.123.073601} is that the nontopologically protected bulk states are also localized at the corner/hinge, which is called skin effect \cite{PhysRevB.97.121401,PhysRevLett.122.076801,PhysRevB.99.201411,PhysRevB.99.081302}. It causes the breakdown of BBC. To solve this problem, the generalized Brillouin zone (GBZ) \cite{PhysRevLett.122.076801,PhysRevB.99.201411} and the biorthogonal expectation value \cite{PhysRevB.99.081302} have been proposed.

Periodic driving has become a versatile tool in artificially creating novel topological phases in systems of ultracold atoms \cite{RevModPhys.89.011004,PhysRevLett.116.205301}, photonics \cite{Rechtsman2013,PhysRevLett.122.173901}, superconductor qubits \cite{Roushan2017}, and graphene \cite{McIver2020}.  Changing symmetry and inducing effective long-range hoppings \cite{PhysRevB.87.201109,PhysRevB.93.184306,PhysRevA.100.023622}, it offers systems an engineered dimension. Parallel to the SOTIs in static systems, the ones in periodic systems are called Floquet SOTIs. Many intriguing features of them absent in static systems have been synthesized by periodic driving in Hermitian systems \cite{PhysRevB.99.045441,PhysRevLett.123.016806,PhysRevB.100.115403,PhysRevLett.124.057001,PhysRevB.101.235403,PhysRevLett.124.216601,PhysRevB.100.085138,PhysRevResearch.1.032045,PhysRevResearch.1.032013}.  To prompt their applications, the creation of exotic non-Hermitian SOTIs in a controllable way via periodic driving is desired. However, the interplay between the skin effect and the periodic driving makes it hard to characterize the Floquet SOTIs. First, it has been revealed in Hermitian systems that Floquet SOTIs are associated with not only the zero-mode corner/hinge states but also the $\pi/T$-mode ones, which are insufficient to be described by the topological invariants in static systems \cite{PhysRevB.99.045441,PhysRevLett.123.016806,PhysRevB.100.115403,PhysRevLett.124.057001,PhysRevB.101.235403,PhysRevLett.124.216601,PhysRevB.100.085138}. Second, the reconstructed symmetries by periodic driving \cite{PhysRevB.87.201109} make it hard to apply the well developed method to recover BBC in static systems \cite{PhysRevLett.122.076801,PhysRevB.99.201411,PhysRevB.99.081302}, where symmetries play an important role \cite{PhysRevLett.123.246801}, to the periodic cases. Floquet SOTIs are studied only in the special case without skin effect \cite{PhysRevB.102.094305}. Thus, a complete description to the non-Hermitian Floquet SOTIs is lacking.

In this Letter, we investigate the Floquet SOTIs in the non-Hermitian systems. The GBZ method developed in static systems is successfully generalized to recover the BBC in our periodically driven systems. For the 2D system, the recovery of the chiral symmetry by the proposed similarity transformations permits us to define a pair of mirror-graded winding numbers to describe zero- and $\pi/T$-mode corner states, whose relationship to the real-space quadrupolar moment is further uncovered. For the 3D system, the Floquet SOTIs are characterized by the mirror Chern number. Exotic non-Hermitian phases of a widely tunable number of 2D corner states and 3D hinge states and a coexistence of first- and second-order TIs, which are hard to present in natural materials, are found in both systems. Filling the blank of theoretical description to the SOTIs of the periodically driven non-Hermitian systems, our results pave the way to apply the non-Hermitian SOTIs via efficiently controlling the numbers of corner/hinger states.

\begin{figure}[tbp]
\centering
\includegraphics[width=0.98\columnwidth]{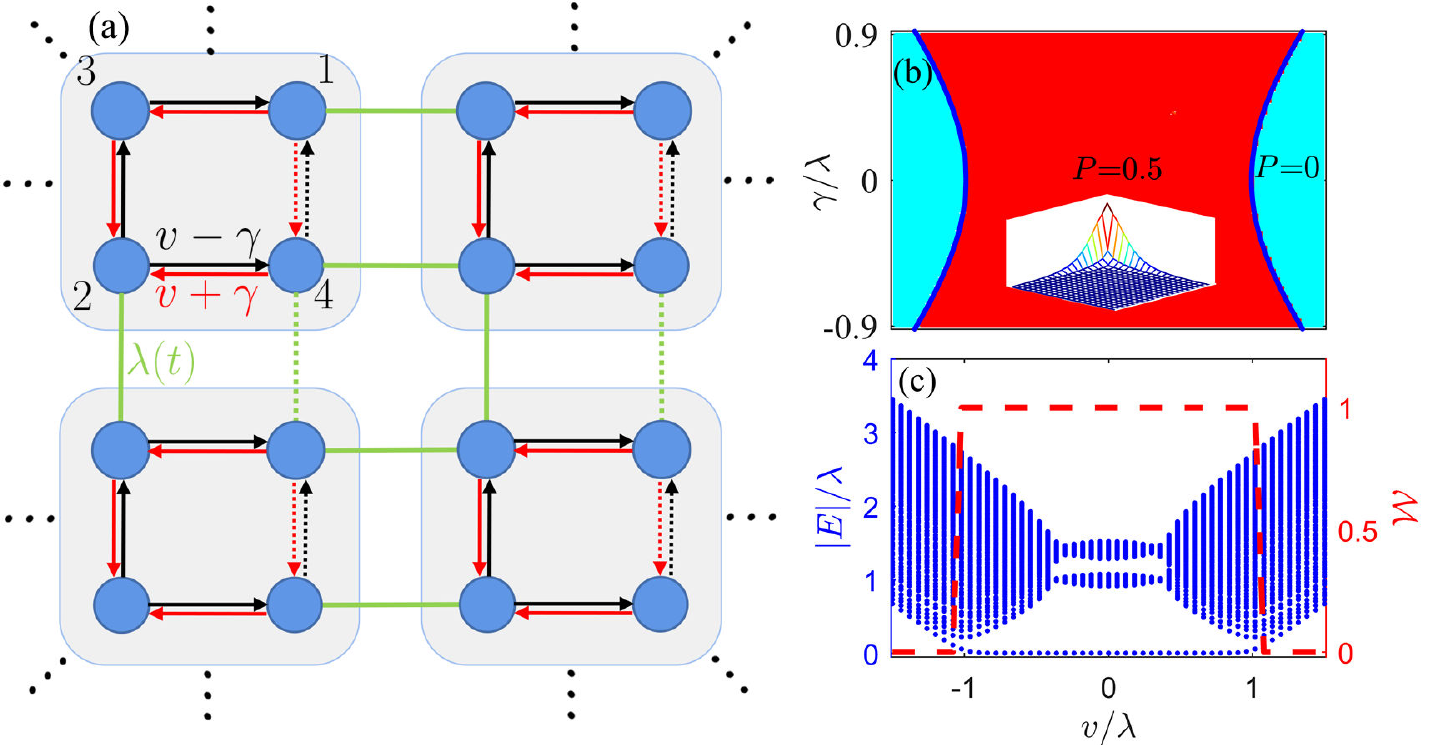}
\caption{(a) Schematic illustration of an $N\times N$ square lattice with an intercell hopping rate $\lambda$ and nonreciprocal intracell ones $v\pm\gamma$. The dashed lines denote the hopping rates with a $\pi$-phase difference from their solid counterparts. (b) Phase diagram characterized by the quadrupole moment $P$ in the absence of the periodic driving. (c) Energy spectrum in the disk geometry and winding number $\mathcal{W}$ when $\gamma=0.4\lambda$. The inset of (b) shows the probability distribution of one of the four zero-mode states when $v=0.8\lambda$. We use $N=20$. } \label{traj}
\end{figure}

\section{2D non-Hermitian system}
We consider a non-Hermitian fermionic system on an $N\times N$ square lattice with nonreciprocal intracell hopping rates [see Fig. \ref{traj}(a)]. Its Hamiltonian reads \cite{PhysRevLett.122.076801}
\begin{eqnarray}
&&H_\text{2D}=\sum_{\mathbf{n}}\{(v-\gamma)[c^{\dag}_{\mathbf{n},1}(c_{\mathbf{n},3}-c_{\mathbf{n},4})+(c^{\dag}_{\mathbf{n},3}+c^{\dag}_{\mathbf{n},4})c_{\mathbf{n},2}]\nonumber\\
  &&~+(v+\gamma)[(c^{\dag}_{\mathbf{n},3}-c^{\dag}_{\mathbf{n},4})c_{\mathbf{n},1}+c^{\dag}_{\mathbf{n},2}(c_{\mathbf{n},4}+c_{\mathbf{n},3})]+\lambda[c^{\dag}_{\mathbf{n},1}\nonumber \\
  &&~\times(c_{\mathbf{n}+\hat{x},3}-c_{\mathbf{n}+\hat{y},4})+c^{\dag}_{\mathbf{n},2}(c_{\mathbf{n}-\hat{y},3}+c_{\mathbf{n}-\hat{x},4})+\rm{H.c.}]\},~~~\label{Hamt}
\end{eqnarray}
where $c^{\dag}_{\mathbf{n},j}$ ($j=1, 2, 3, 4$) is the fermionic creation operator at the $j$th sublattice of unit-cell site $\mathbf{n}=(n_x,n_y)$, $\hat{x}$ and $\hat{y}$ are the unit vectors of the $x$ and $y$ directions, $\lambda$ and $v\pm \gamma$ are the intercell and the nonreciprocal intracell hopping rates. The SOTI is characterized by the quadrupole moment. Inspired by its definition in Hermitian systems \cite{PhysRevB.100.245135,PhysRevB.100.245134} and the biorthogonal basis in non-Hermitian systems \cite{PhysRevLett.123.246801}, we construct a non-Hermitian quadrupole moment in the disk geometry, i.e., the open boundary condition in both directions \cite{SMP}. Figure \ref{traj}(b) shows the phase diagram characterized by $P$. It reveals a phase transition at $|v|=\sqrt{\gamma^2+\lambda^2}$. When $|v|<\sqrt{\gamma^2+\lambda^2}$, $P=0.5$ signifies the formation of a SOTI. The real-space energy spectrum in Fig. \ref{traj}(c) confirms the presence of a four-fold degenerate zero-mode state, which distributes at the corner [see the inset of Fig. \ref{traj}(b)].

The corner nature of the zero-mode state is guaranteed by the symmetries of the system. Under the periodic boundary condition in $x$ and $y$ directions, we have $H=\sum_{\bf k}{\bf C}_{\bf k}^\dag\mathcal{H}_\text{2D}({\bf k}){\bf C}_{\bf k}$ with ${\bf C}^\dag_{\bf k}=( \begin{array}{cccc} c_{{\bf k},1}^\dag & c_{{\bf k},2}^\dag  & c_{{\bf k},3}^\dag  & c_{{\bf k},4}^\dag  \\\end{array})$ and
\begin{equation}
\begin{split}
\mathcal H_\text{2D}({\bf k})&=(v+\lambda \cos k_x)\tau_x\sigma_0-(\lambda\sin k_x+i\gamma)\tau_y\sigma_z\\
    &+(v+\lambda \cos k_y)\tau_y\sigma_y+(\lambda \sin k_y+i\gamma)\tau_y\sigma_x
\end{split},\label{model}
\end{equation}
where $\tau_i$ and $\sigma_i$ are Pauli matrices, and $\tau_0$ and $\sigma_0$ are identity matrices. Equation \eqref{model} possesses the mirror-rotation symmetry $\mathcal{M}_{xy}\mathcal{H}(k_x,k_y)\mathcal{M}^{-1}_{xy}=\mathcal{H}(k_y,k_x)$ with $\mathcal{M}_{xy}=[(\tau_0-\tau_z)\sigma_x-(\tau_0+\tau_z)\sigma_z]/2$ and the chiral symmetry $\mathcal{S}\mathcal{H}({\bf k})\mathcal{S}^{-1}=-\mathcal{H}({\bf k})$ with $\mathcal{S}=\tau_z\sigma_0$ \cite{PhysRevLett.122.076801}. Thus, its upper and lower bands are two-fold degenerate. The bands coalesce at the exceptional points when $|v|=|\lambda\pm\gamma|$, which do not match with the critical points of phase transition obtained under the open boundary condition. It is called the non-Hermiticity induced breakdown of the BBC \cite{PhysRevLett.121.086803}.

The breakdown causes that we cannot use the well-defined Bloch band theory in Hermitian systems to characterize the non-Hermitian SOTIs. A non-Bloch band theory has been developed to recover the BBC in $\mathcal{H}(\tilde{\bf k})$ by introducing the GBZ $\tilde{\bf k}\equiv(k_x-i \ln r,k_y-i\ln r)$ with $r=[|(v-\gamma)/(v+\gamma)|]^{1/2}$ \cite{SMP}. Due to the mirror-rotation symmetry, the SOTIs are sufficiently described by $\mathcal{H}(\tilde{k},\tilde{k})$ along the high-symmetry line $\tilde{k}_x=\tilde{k}_y\equiv\tilde{k}$ \cite{PhysRevLett.122.076801}. Diagonalizing $\mathcal{H}(\tilde{k},\tilde{k})$ into $\text{diag}[\mathcal{H}^+(\tilde{k}),\mathcal{H}^-(\tilde{k})]$ with $\mathcal{H}^\alpha(\tilde{k})={\bf h}^\alpha(\tilde{k})\cdot{\pmb\sigma}$ and $\alpha=\pm$, we can describe its bulk topology by the mirror-graded winding number $\mathcal{W}=(\mathcal{W}_{+}-\mathcal{W}_{-})/2$. Here we define
$\mathcal{W}_\alpha=\frac{i}{2\pi}\int_{0}^{4\pi}\frac{\langle u^\text{L}_{\alpha}\lvert\partial_k\lvert u^\text{R}_{\alpha}\rangle}{\langle u^\text{L}_{\alpha}\lvert u^\text{R}_{\alpha}\rangle}dk$  \cite{SMP}, where $\lvert u^\text{L,R}_{\alpha}\rangle$ are the right and left eigenstates of $\mathcal{H}^\alpha(\tilde{k})$ \cite{PhysRevLett.122.076801}. When $|v|<\sqrt{\gamma^2+\lambda^2}$, the system has $\mathcal{W}=1$ and hosts four degenerate corner states [see Fig. \ref{traj}(c)]. It means that the quadrupole moment and the mirror-graded winding number can describe the static system equivalently.

\section{2D Floquet SOTIs}
Consider that the intercell hopping rate is periodically driven between two specific $\lambda_1$ and $\lambda_2$ in the respective time duration $T_1$ and $T_2$
\begin{equation}
\lambda(t) =\begin{cases}\lambda_1=q_1 f , ~t\in\lbrack mT, mT+T_1)\\\lambda_2=q_2f, ~t\in\lbrack mT+T_1, (m+1)T)\end{cases}, \label{procotol}
\end{equation}
where $m\in \mathbb{Z}$, $T=T_1+T_2$ is the driving period, and $\lambda_j$ are nondimensionalized into $q_j$. The system ${H}(t)= {H}(t+T)$ has no well-defined energy spectrum because the energy is not conserved. According to Floquet theorem, the one-period evolution operator $U_T=e^{-iH_2T_2}e^{-iH_1T_1}$ defines an effective Hamiltonian ${H}_\text{eff}=i\ln U_T/T$ whose eigenvalues are called quasienergies \cite{PhysRevA.7.2203}. The topological properties of our periodic system are defined in such a quasienergy spectrum \cite{PhysRevB.87.201109}. We can identify the Floquet SOTIs by calculating the non-Hermitian quadrupole moment from the real-space ${H}_\text{eff}$. The real part of quasienergy is a phase factor defined modulus $2\pi/T$, which leads to that the edge states in the periodic system can occur not only in the real part of quasienergy zero but also $\pi/T$.

To establish the topological description to the Floquet SOTIs, we apply Floquet theorem to Eq. \eqref{model} in the momentum space and obtain $\mathcal{H}_\text{eff}({\bf k})$. It inherits the mirror-rotation symmetry, which suffices us to study the high-symmetry-line $\mathcal{H}_\text{eff}(k,k)=\text{diag}[\mathcal{H}^+_\text{eff}(k),\mathcal{H}^-_\text{eff}(k)]$ with $\mathcal{H}_\text{eff}^\alpha(k)=i\ln[e^{-i\mathcal{H}^\alpha_2(k)T_2}e^{-i\mathcal{H}^\alpha_1(k)T_1}]/T$. First, we can obtain from $\mathcal{H}_\text{eff}(k,k)$ that the phase transitions occur for ${k}$ and the parameters satisfying either \cite{SMP}
\begin{eqnarray}
&&T_jE^{\alpha}_j=n_j\pi,~n_j\in \mathbb{Z}, \label{gen}\\
\text{or}~&&
\begin{cases}
\underline{\mathbf{h}}^{\alpha}_1\cdot\underline{\mathbf{h}}^{\alpha}_2=\pm1\\
T_1{E}^{\alpha}_1\pm T_2{E}^{\alpha}_2=n\pi,~n\in\mathbb{Z}
\end{cases}\label{hh1}
\end{eqnarray}
at the quasienergy zero (or $\pi/T$) if $n$ is even (or odd), where $\underline{\mathbf{h}}^\alpha_j=\mathbf{h}^\alpha_j/E^\alpha_j$ and $E^\alpha_j=(\mathbf{h}^\alpha_j\cdot\mathbf{h}^\alpha_j)^{1/2}$ are the complex eigen energies of $\mathcal{H}^\alpha_j(k)$ \cite{SMP}. As the sufficient condition for the topological phase transition, Eqs. \eqref{gen} and \eqref{hh1} supply a guideline to manipulate the driving parameters for Floquet engineering to various novel topological phases at will. Second, we can define proper topological invariants from $\mathcal{H}_\text{eff}(k,k)$ to characterize the zero- and $\pi/T$-mode corner states. This is nontrivial because the chiral symmetry $\mathcal{S}$ is broken in $\mathcal{H}_\text{eff}(k,k)$ due to $[\mathcal{H}_1(k,k),\mathcal{H}_2(k,k)]\neq0$. We propose the following scheme to recover the symmetry. Two similarity transformations $G_j=e^{i(-1)^{j}\mathcal{H}_j(k,k)T_j/2}$ ($j=1,2$) convert $U_T$ into $\tilde{U}_{1}$ and $\tilde{U}_{2}$. The effective Hamiltonians defined in $\tilde{U}_j$ via $\tilde{\mathcal H}_{\text{eff},j}(k,k)\equiv i\ln \tilde{U}_j/T$ recover the chiral symmetry \cite{PhysRevB.102.041119}. Then after introducing the GBZ, the winding numbers $\mathcal{W}_j$ defined in $\tilde{\mathcal H}_{\text{eff},j}(\tilde{k},\tilde{k})$ can topologically characterize the Floquet SOTIs. The numbers of the zero-mode and $\pi/T$-mode corner states relate to $\mathcal{W}_j$ as \cite{PhysRevB.90.125143}
\begin{equation}
N_0=2|\mathcal{W}_1+\mathcal{W}_2|,~N_{\pi/T}=2|\mathcal{W}_1-\mathcal{W}_2|. \label{tpnb}
\end{equation}
As we will confirm later, the quadrupolar moment relates to the numbers $N_0$ and $N_{\pi/T}$ of the corner states as
\begin{equation}
P={1\over2}\left[(|N_0|+|N_{\pi/T}|)\,\,\text{mod}\,\,2\right].  \label{preln}
\end{equation}It has been used to define the $Z_2$ index for quantum spin Hall systems \cite{PhysRevB.79.195321}.

\section{Numerical results}
\begin{figure}[tbp]\centering
\includegraphics[width=1\columnwidth]{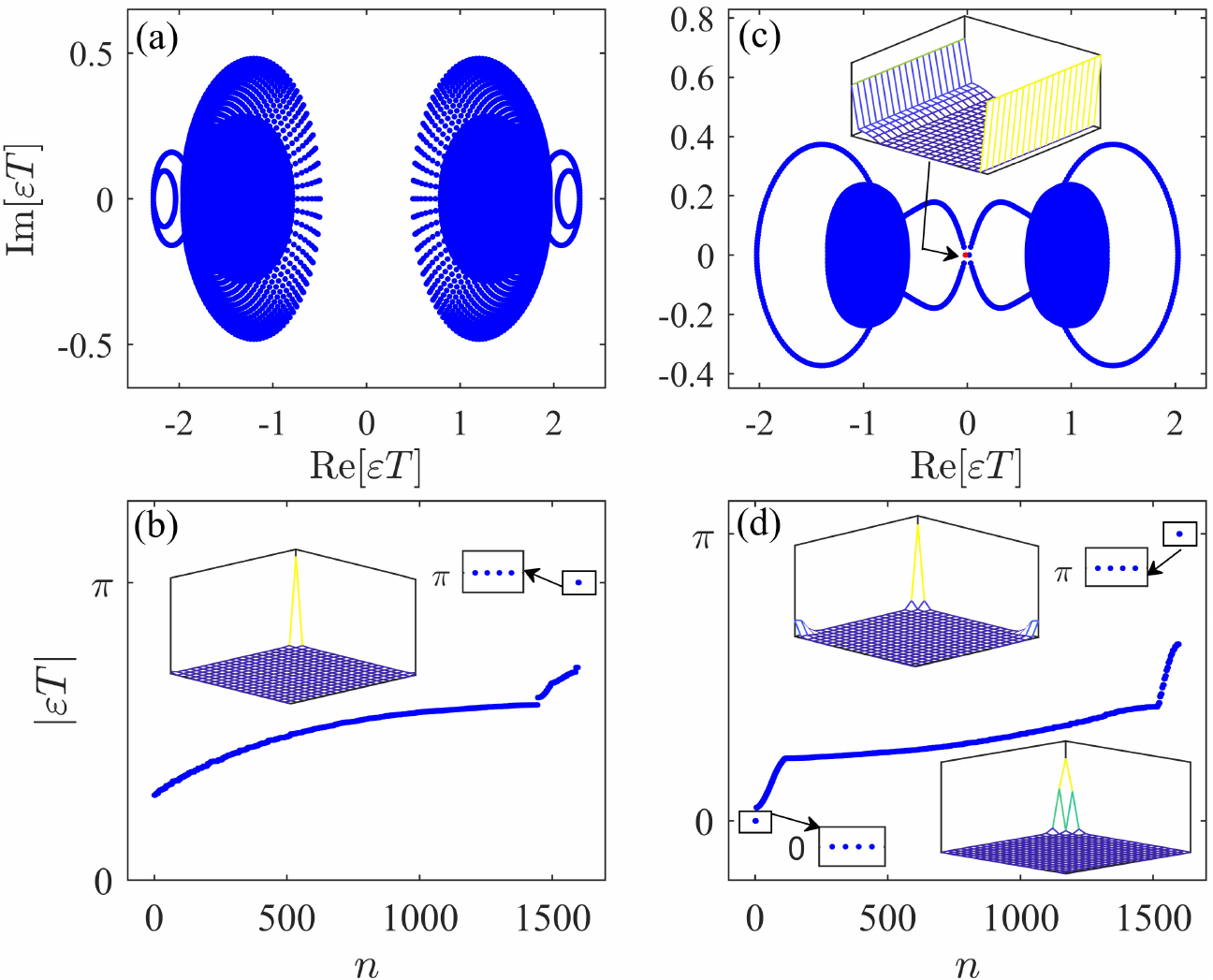}
\caption{Quasienergies in the cylinder [(a),(c)] and disk [(b),(d)] geometries when $T_1=T_2=1.6f^{-1}$ in [(a),(b)] and $2.2f^{-1}$ in [(c),(d)]. The probability distributions of the edge/corner states are given in the insets. We use $v=0.8f$, $\gamma=0.4f$, $q_1=1$, $q_2=0$, and $N=20$. }\label{engsp}
\end{figure}

To reveal the distinguished role of the periodic driving in inducing novel SOTIs, we plot in Fig. \ref{engsp} the quasienergies in cylinder (i.e., open boundary condition only in the $x$ direction) and disk geometries. The quasienergies in cylinder geometry in Fig. \ref{engsp}(a) show a gapped phase, while the corresponding ones in disk geometry in Fig. \ref{engsp}(b) show the formation of four degenerate $\pi/T$-mode corner states. It reveals a Floquet topological phase trivial in the first order but nontrivial in the second order, which is similar to the conventional SOTIs \cite{Benalcazar61,Schindlereaat0346}. Figures \ref{engsp}(c) and \ref{engsp}(d) show an anomalous phase which supports not only a SOTI in disk geometry but also a first-order topological insulator in cylinder geometry. Such coexistence of the first- and second-order topological phase is absent in its original static Hamiltonian. It indicates that the periodic driving, as a useful tool in controlling phase transition, enables us to obtain exotic phases completely absent in its static counterpart.

\begin{figure}[tbp]
\centering
\includegraphics[width=1\columnwidth]{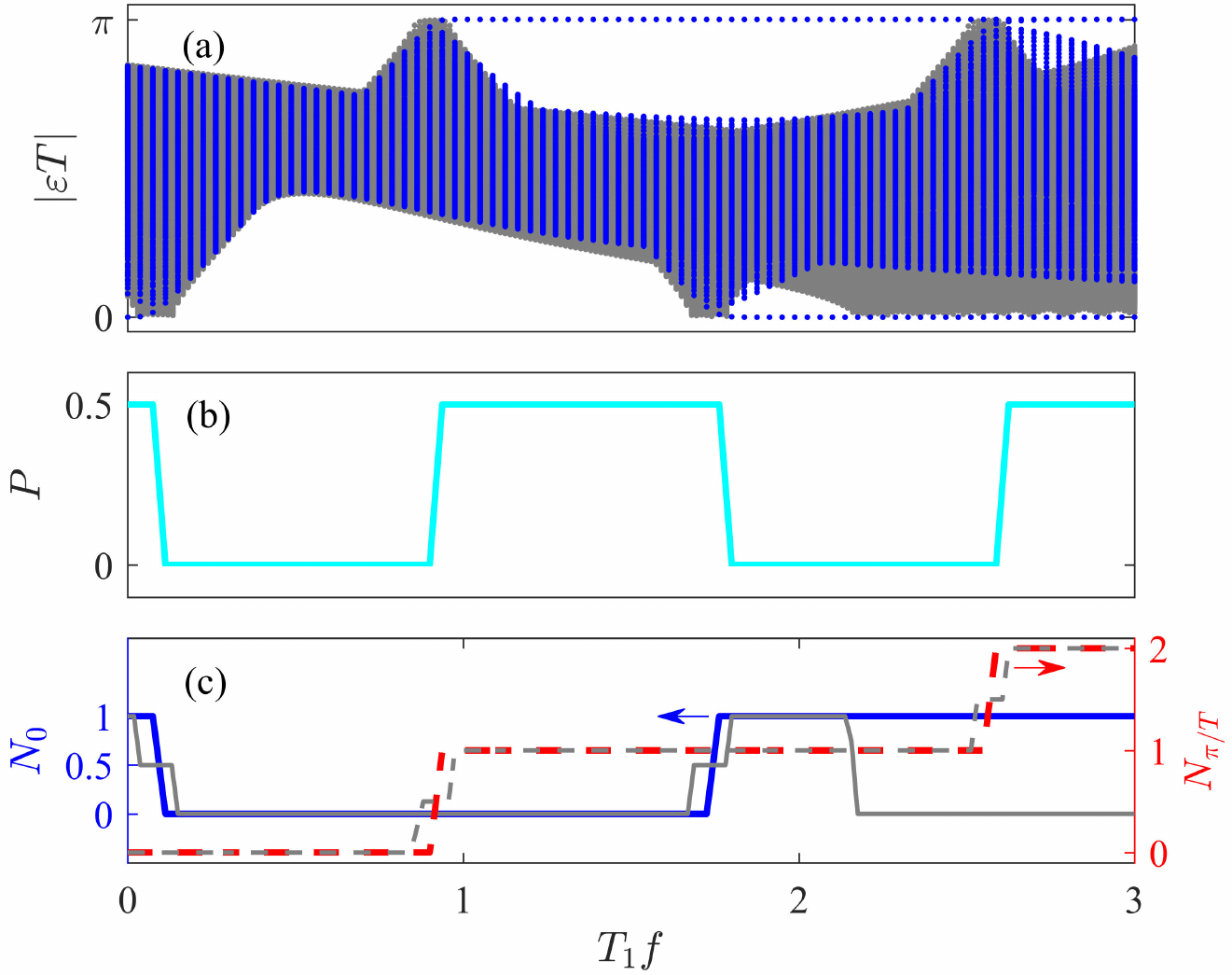}
\caption{Quasienergy spectra from the open (blue area) and periodic (gray area) boundary conditions (a), quadrupole moment (b), and numbers of the zero-mode [blue solid line in (c)] and $\pi/T$-mode [red dashed line in (c)] corner states from the GBZ and the conventional BZ [gray lines in (c)] as the change of $T_1$. We use $T_2=0.7f^{-1},v=1.2f$, $\gamma=0.2f$, $N=30$, and $q_1=-q_2=1.5f$.} \label{pad}
\end{figure}

Figure \ref{pad}(a) shows the quasienergy spectrum in the disk geometry, which has a dramatic difference from the one in the momentum space [see the gray area of Fig. \ref{pad}(a)]. It is the breakdown of the BBC. After mapping the conventional Brillouin zone (BZ) to the GBZ, we can retrieve the BBC and obtain the band-closing points from Eqs. \eqref{gen} and \eqref{hh1}. Because the eigenvalues of $\mathcal{H}^\pm_\text{eff}(\tilde{k})$ are degenerate, it suffices to evaluate one of them. Using ${\bf h}^+_j(\tilde{k})=\sqrt{2}(v+\lambda_j\cos\tilde{k},i\gamma+\lambda_j\sin\tilde{k},0)$ and focusing on $v>\gamma>0$, we have the band-closing conditions as follows.

\textbf{Case I:} Equation \eqref{gen} results in that any $k$ satisfying
\begin{equation}
[2(\kappa^2+\lambda^2_j+2\lambda_j\kappa\cos k)]^{1/2}T_j=n_j\pi\label{b1}
\end{equation}with $\kappa=(v^2-\gamma^2)^{1/2}$ gives a phase transition.

\textbf{Case II:} $\underline{\mathbf{h}}_1\cdot\underline{\mathbf{h}}_2=\pm1$ requires $k$ in $\tilde{k}$ to be $\theta=0$ or $\pi$. According to Eq. \eqref{hh1}, we obtain that
\begin{equation}
\sqrt{2}(|\kappa+\lambda_1e^{i\theta}|T_1\pm|\kappa+\lambda_2e^{i\theta}|T_2)=n_{\theta,\pm}\pi,\label{b2}
\end{equation}
for $\text{sgn}[\prod_{j=1}^2(\kappa+e^{i\theta}\lambda_j)]=\pm1$ gives a phase transition.
\\It can be verified that the band closing points at $T_1=0.08f$, $0.91f$, $1.74f$, and $2.57f$ in Fig. \ref{pad}(a) can be analytically reproduced by Eq. \eqref{b2} with $n_{0,-}=0$, $1$, $2$, and $3$. The Floquet SOTIs can be captured by the real-space quadrupole moment [see Fig. \ref{pad}(b)]. However, as a $Z_2$ topological number, it characterizes only the parity not the explicit value of the number of the corner states. After recovering the chiral symmetry in $\tilde{\mathcal H}_{\text{eff},j}$ by the similarity transformation $G_j$, we can define winding numbers $\mathcal{W}_j$ in $\tilde{\mathcal H}_{\text{eff},j}$ to realize this. The gray lines in Fig. \ref{pad}(c) show the number of corner states by calculating $\mathcal{W}_j$ in the conventional BZ. Although reflecting the band-closing behavior in the momentum space, the ill-defined topological numbers nonphysically take half integers \cite{PhysRevLett.116.133903} and are irrelevant to the corner states. However, the ones from the GBZ, which relate to $P$ in Fig. \ref{pad}(b) according to Eq. \eqref{preln}, correctly count the numbers of the corner states [see the blue solid and red dashed lines in Fig. \ref{pad}(c)].

\begin{figure}[tbp]
\centering
\includegraphics[width=1\columnwidth]{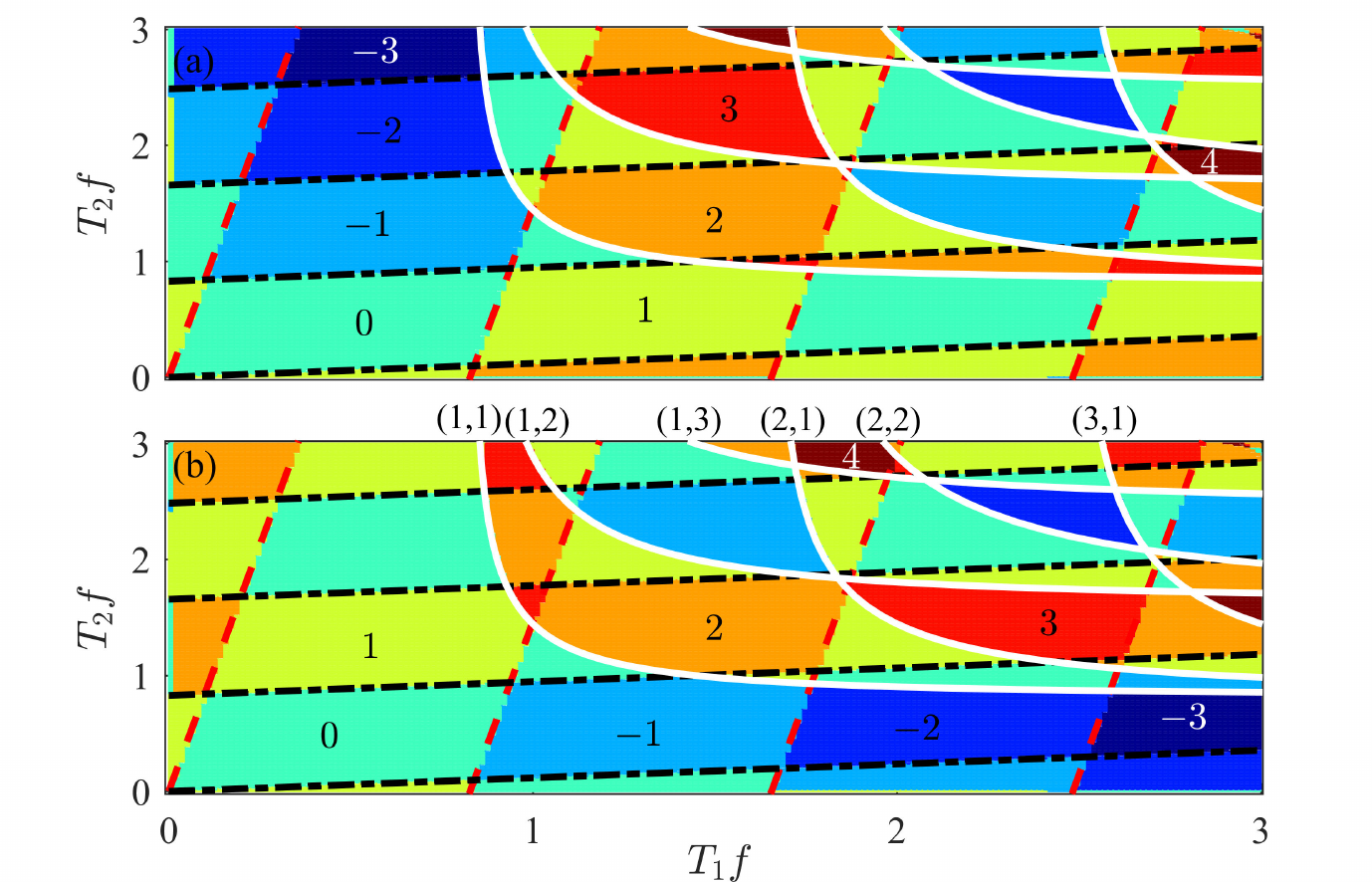}
\caption{Phase diagram characterized by $\mathcal{W}_1$ (a) and $\mathcal{W}_2$ (b). The white solid lines are from Eqs. \eqref{b1} with $(n_1,n_2)$ labeled explicitly. The red dashed and the black dot-dashed lines are from Eqs. \eqref{b2} with $n_{0,-}=0,1,2,3$ and $n_{\pi,-}=0,-1,-2,-3$, respectively. The other parameters are the same as Fig. \ref{pad}.} \label{dsord}
\end{figure}

To get a global understanding on the Floquet SOTIs, we plot in Fig. \ref{dsord} the phase diagram in the $T_1$-$T_2$ plane. It is remarkable to see that Floquet SOTIs with widely tunable numbers of corner states, which are much richer than the static case, are induced. Different types of phase boundaries are all analytically explainable by Eqs. \eqref{b1} and \eqref{b2} (see the caption of Fig. \ref{dsord}). This verifies the validity of the description of non-Hermitian Floquet SOTIs we developed. The result implies that we can manipulate the corner states at ease and on demand by the periodic driving. Being hard to realize in natural materials, such controllability favored by the periodic driving might inspire insightful application of the non-Hermitian SOTIs.

\begin{figure}[tbp]
\centering
\includegraphics[width=1\columnwidth]{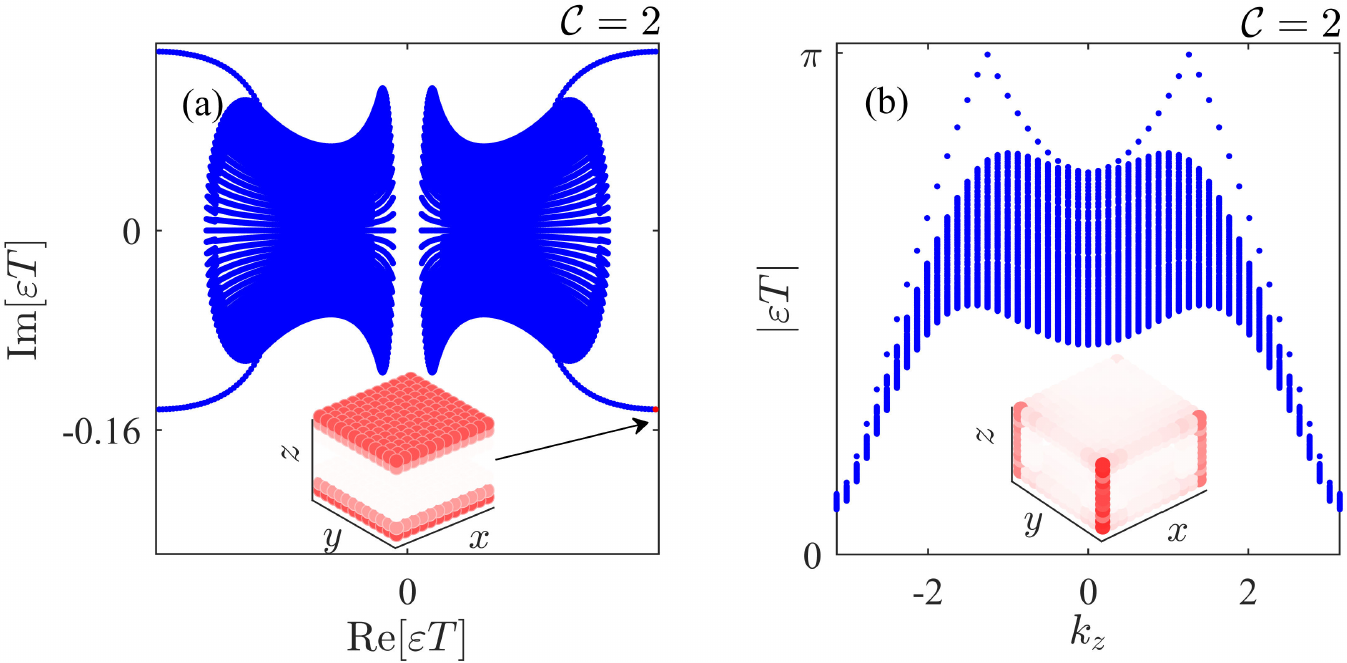}
\caption{ Quasienergy spectra under the $z$-direction open boundary condition with $k_x=k_y$ in (a) and the $x, y$-direction open boundary condition in (b). The probability distributions of the corresponding surface and hinge states are shown in the insets. We use $v=1.3f$, $\gamma=0.2f$, $\chi=\chi'=f$, $q_1=-q_2=1.5$, $T_1=T_2=0.6f^{-1}$, and $N=15$.  } \label{bce}
\end{figure}

\section{3D Floquet SOTI}
Generalizing Eq. \eqref{Hamt} to the 3D layered structure by further considering the intracell hopping between the two neighboring layers in Fig. \ref{traj}(a), we have its momentum-space Hamiltonian
\begin{eqnarray}
\mathcal{H}_\text{3D}({\bf k})&=&\mathcal H_\text{2D}({\bf k})+\chi\cos k_z(\tau_x\sigma_0+\tau_y\sigma_y)\nonumber\\&&+\sqrt{2}\chi'\sin k_z\tau_z\sigma_0,\label{3dH}
\end{eqnarray}where $\chi$ and $\chi'$ are the intracell hopping rates between different and same sublattices of the two layers, respectively.
The mirror-rotation symmetry $\mathcal{M}_{xy}$ is respected. Thus the bulk topology is characterized by the high-symmetry-line $\mathcal{H}(k,k,k_z)$, which can be diagonalized into diag[$\mathcal{H}^{+}(k,k_z),\mathcal{H}^{-}(k,k_z)$]. Its bulk topology is described by the mirror Chern number $\mathcal{C}=(\mathcal{C}_{+}-\mathcal{C}_{-})/2$ with
$\mathcal{C}_{\alpha}=\frac{1}{4\pi}\int_\text{BZ}\frac{1}{E^{3/2}}\mathbf{h}^{\alpha}\cdot(\partial_{k}\mathbf{h}^{\alpha}\times\partial_{k_z}\mathbf{h}^{\alpha})d^2\mathbf{k}$ \cite{Schindlereaat0346}. For the static system, $\mathcal{C}$ only can be $\pm1$.

Besides the high controllability of hinge states to the 3D SOTIs, an exotic consequence delivered by the periodic driving \eqref{procotol} to Eq. \eqref{3dH} is the coexistence of the first- and second-order topological phases. The quasienergy spectrum under the $z$-direction open boundary condition in Fig. \ref{bce}(a) shows two chiral surface modes. This is a first-order topological phase with $\mathcal{C}=2$, which is calculated from the traditional BZ due to the absence of skin effect in the $z$ direction. The corresponding quasienergy spectrum under the $x,y$-direction open boundary condition in Fig. \ref{bce}(b) hosts two chiral boundary modes, which corresponds to two four-fold degenerate hinge states under the three-direction open boundary condition [see the inset of Fig. \ref{bce}(b)]. It signifies the formation of a 3D SOTI. After introducing the GBZ to recover the BBC, we obtain that the SOTI can be well described by its bulk topological number $\mathcal{C}=2$. The result confirms the coexistence of the first- and second-order Floquet topological phases in the 3D non-Hermitian system. It has been recently found that such a phase can generate exotic higher-order Weyl semimetal \cite{PhysRevLett.125.146401}.

\section{Conclusion}
In summary, we have investigated the SOTIs in 2D and 3D periodically driven non-Hermitian systems. The complete description to the non-Hermitian Floquet SOTIs is established by introducing the GBZ to retrieve the BBC. Diverse exotic non-Hermitian topological phases of widely tunable numbers of 2D corner states and 3D hinge states and coexistence of the first- and second-order topological phases are induced by the periodic driving. For the 2D case, the relationship between the non-Hermitian quadrupolar moment and the mirror-graded winding numbers has been uncovered. Our result opens an avenue to artificially generate the exotic non-Hermitian SOTIs absent in natural material, which is useful in exploring their application.  The observation of the SOTIs in the static case \cite{Stefan2018,PhysRevB.102.104109,Mittal2019} and the Floquet topological phases in the non-Hermitian electric and optical systems \cite{Helbig2020,Weidemann311} indicates that our result is realizable in the recent experimental state of the art.

\section{Acknowledgments}
The work is supported by the National Natural Science Foundation (Grant Nos. 11875150 and 11834005).
\bibliography{references}

\end{document}

\title{Supplemental material for ``Floquet second-order topological insulators in non-Hermitian systems''}
\author{Hong Wu}
\affiliation{School of Physical Science and Technology, Lanzhou University, Lanzhou 730000, China}
\author{Bao-Qin Wang}
\affiliation{School of Physical Science and Technology, Lanzhou University, Lanzhou 730000, China}
\author{Jun-Hong An}
\email{anjhong@lzu.edu.cn}
\affiliation{School of Physical Science and Technology, Lanzhou University, Lanzhou 730000, China}

\maketitle

\section{Quadrupole moment}
The SOTI is characterized by the quadrupole moment. Inspired by its definition in Hermitian systems \cite{PhysRevB.100.245134, PhysRevB.100.245135} and the biorthogonal basis in non-Hermitian systems \cite{PhysRevLett.123.246801}, we construct a non-Hermitian quadrupole moment in the real-space disk geometry, i.e., the open boundary condition in both directions, as
\begin{equation}
P=\Big[\frac{\text{Im}\ln\det\mathcal{U}}{2\pi}-\sum_{{\bf n},i;{\bf m},j} \frac{X_{{\bf n},i;{\bf m},j}}{2L_xL_y}\Big]\text{mod}~1.\label{polrz}
\end{equation}
Here the elements of $\mathcal{U}$ read $\mathcal{U}_{\alpha\beta}\equiv \langle \alpha^\text{L}|e^{i2\pi X/(L_xL_y)}|\beta^\text{R}\rangle$ with $H|\alpha^\text{R} \rangle=E_{\alpha}|\alpha^\text{R}\rangle$ and $H^{\dag}|\alpha^\text{L} \rangle=E^{*}_{\alpha}|\alpha^\text{L}\rangle$, and the coordinate $X_{{\bf n},i;{\bf m},j}=n_xn_y\delta_{{\bf n}{\bf m}}\delta_{ij}$ with $i,j=1,\cdots,4$ being the sublattices. In the Hermitian case, Eq. \eqref{polrz} returns to the form in Refs. \cite{PhysRevB.100.245134, PhysRevB.100.245135}.

\section{Generalized Brillouin zone}
In this section, we use the method developed in the 1D Su-Schrieffer-Heeger model \cite{PhysRevLett.121.086803} to calculate the generalized Brillouin zone of our system. The skin effect can be removed by a similarity transformation in the real space
\begin{eqnarray}
S&=&\exp\Big[\sum_{{\bf n}}(c^{\dag}_{\mathbf{n},1}c_{\mathbf{n},1}+r^{-2}c^{\dag}_{\mathbf{n},2}c_{\mathbf{n},2})\ln r^{n_x+n_y-1}\nonumber\\
&&+(c^{\dag}_{\mathbf{n},3}c_{\mathbf{n},3}+c^{\dag}_{\mathbf{n},4}c_{\mathbf{n},4})\ln r^{n_x+n_y-2}\Big].
\end{eqnarray}
It converts Eq. (1) in the main text into $\bar{H}=S^{-1}H_\text{2D}S$ with
\begin{eqnarray}
\bar{H}&&=\sum_{\mathbf{n}}\{r^{-1}(v-\gamma)[c^{\dag}_{\mathbf{n},1}(c_{\mathbf{n},3}-c_{\mathbf{n},4})+(c^{\dag}_{\mathbf{n},3}+c^{\dag}_{\mathbf{n},4})c_{\mathbf{n},2}]\nonumber\\
  &&+r(v+\gamma)[(c^{\dag}_{\mathbf{n},3}-c^{\dag}_{\mathbf{n},4})c_{\mathbf{n},1}+c^{\dag}_{\mathbf{n},2}(c_{\mathbf{n},4}+c_{\mathbf{n},3})]+\lambda[c^{\dag}_{\mathbf{n},1}\nonumber \\
  &&\times(c_{\mathbf{n}+\hat{x},3}-c_{\mathbf{n}+\hat{y},4})+c^{\dag}_{\mathbf{n},2}(c_{\mathbf{n}-\hat{y},3}+c_{\mathbf{n}-\hat{x},4})+\rm{H.c.}]\}.~~~~~\label{Hamtsm}
\end{eqnarray}
One can readily check that, as long as $r=\sqrt{(v-\gamma)/(v+\gamma)}$, all the bulk states of $\bar{H}$ do not reside in the edges anymore and thus the skin effect is removed \cite{PhysRevLett.122.076801}. Rewriting Eqs. \eqref{Hamtsm} in the momentum space, we have $\bar{H}=\sum_{\bf k}{\bf C}_{\bf k}^\dag\mathcal{\bar H}_\text{2D}({\bf k}){\bf C}_{\bf k}$ with ${\bf C}^\dag_{\bf k}=( \begin{array}{cccc} c_{{\bf k},1}^\dag & c_{{\bf k},2}^\dag  & c_{{\bf k},3}^\dag  & c_{{\bf k},4}^\dag  \\\end{array})$, $\mathcal{\bar H}_\text{2D}({\bf k})=\left(
\begin{array}{cc}
 0  &  \mathcal{\bar H}_1 \\
\mathcal{\bar H}_2  &   0
\end{array}
\right)$, and
\begin{eqnarray}
\mathcal{\bar H}_1&=&
\left[
\begin{array}{cc}
 r^{-1}(v-\gamma+\lambda e^{i\tilde{k}_x})   &  -r^{-1}(v-\gamma+\lambda e^{i\tilde{k}_y}) \\
 r(v+\gamma+\lambda e^{-i\tilde{k}_y})   &    r(v+\gamma+\lambda e^{-i\tilde{k}_x} )
\end{array}
\right],~~~~\\
\mathcal{\bar H}_2&=&
\left[
\begin{array}{cc}
 r(v+\gamma+\lambda e^{-i\tilde{k}_x})   &  r^{-1}(v-\gamma+\lambda e^{i\tilde{k}_y}) \\
 -r(v+\gamma+\lambda e^{-i\tilde{k}_y})   &    r^{-1}(v-\gamma+\lambda e^{i\tilde{k}_x})
\end{array}
\right].~~~
\end{eqnarray}
where $\tilde{ k}_x=k_x-i\ln r$ and $\tilde{ k}_y=k_y-i\ln r$. Comparing $\mathcal{\bar H}_\text{2D}({\bf k})$ with Eq. (2) in the main text, we see that, except for the unimportant pre-factors $r^{\pm1}$, which do not affect the eigenvalues, they have the same form after replacing ${\bf k}$ with $\tilde{\bf k}$. This gives us an insight to remove the skin effect in the original non-Hermitian system by directly replacing ${\bf k}$ in Eq. (2) of the main text by $\tilde{\bf k}$ \cite{PhysRevLett.121.086803}. Such complex vector $\tilde{\bf k}$ forms the a generalized Brillouin zone.

We can see that the similarity transformation $S$ to remove the skin effect is independent on $\lambda$. When the periodic driving
\begin{equation}
\lambda(t) =\begin{cases} \lambda_1=q_1\,f ,& t\in\lbrack mT, mT+T_1)\\ \lambda_2=q_2\,f,& t\in\lbrack mT+T_1, (m+1)T), \end{cases}\label{qd}
\end{equation}
is applied on $\lambda$, we readily have
\begin{equation}
\bar{U}_{T}=S^{-1}U_{T}S=e^{-i\bar{H}_2T_2}e^{-i\bar{H}_1T_1},
\end{equation}
where the $\bar{H}_{j}$ are Eq. \eqref{Hamtsm} with $\lambda=q_jf$. Since no skin effect is present in $\bar{H}_{j}$, neither to $\bar{H}_\text{eff}\equiv {i\over T}\ln \bar{U}_T$. We have successfully removed the skin effect by the same generalized Brillouin zone defined by $S$ as the static case \cite{PhysRevLett.122.076801}. This lays the foundation for studying the topological property of our periodically driven non-Hermitian system.

\section{Mirror-rotation symmetry under periodic driving}
Our periodically driven system inherits the mirror-rotation symmetry of the static system. This can be proven as follows. The one-periodic evolution operator in momentum space reads
\begin{equation}
{U}_T(k_x,k_y)=e^{-i\mathcal{H}_2(k_x,k_y)T_2}e^{-i\mathcal{H}_1(k_x,k_y)T_1}.\label{ssuu}
\end{equation}
Since the static model has the mirror-rotation symmetry $\mathcal{M}_{xy}$, we have
\begin{equation}
\mathcal{M}_{xy}\mathcal{H}_j(k_x,k_y)\mathcal{M}_{xy}^{-1}=\mathcal{H}_j(k_y,k_x).\label{mry}
\end{equation}
It can be proven that
\begin{equation}
\mathcal{M}_{xy}{U}_T(k_x,k_y)\mathcal{M}_{xy}^{-1}=e^{-i\mathcal{H}_2(k_y,k_x)T_2}e^{-i\mathcal{H}_1(k_y,k_x)T_1}.\label{smmuy}
\end{equation}
The combination of Eqs. \eqref{ssuu} and \eqref{smmuy} readily leads to
\begin{equation}
\mathcal{M}_{xy}{U}_T(k_x,k_y)\mathcal{M}_{xy}^{-1}={U}_T(k_y,k_x).
\end{equation}
According to the definition of the effective Hamiltonian $\mathcal{H}_\text{eff}(k_x,k_y)={i\over T}\ln {U}_T(k_x,k_y)$, we obtain
\begin{equation}
\mathcal{M}_{xy}\mathcal{H}_\text{eff}(k_x,k_y)\mathcal{M}_{xy}^{-1}=\mathcal{H}_\text{eff}(k_y,k_x).\end{equation}
Therefore, the effective Hamiltonian inherits the mirror-rotation symmetry of the static system.

\section{Effective Hamiltonian and band-closing points}
Applying Floquet theorem to momentum-space Hamiltonian, we obtain $\mathcal{H}_\text{eff}({\bf k})$. The mirror-rotation symmetry of $\mathcal{H}_\text{eff}({\bf k})$ suffices us to study the high-symmetry-line $\mathcal{H}_\text{eff}(k,k)$. The block-diagonal $\mathcal{H}_j(k,k)=\text{diag}[\mathcal{H}_j^+(k),\mathcal{H}_j^-(k)]$ leads to $\mathcal{H}_\text{eff}(k,k)=\text{diag}[\mathcal{H}_\text{eff}^+(k),\mathcal{H}_\text{eff}^-(k)]$ with $\mathcal{H}_\text{eff}^\alpha(k)={i\over T}\ln[e^{-i\mathcal{H}^\alpha_2(k)T_2}e^{-i\mathcal{H}^\alpha_1(k)T_1}]\equiv {\bf h}_\text{eff}^\alpha\cdot{\pmb\sigma}$ and $\alpha=\pm$. Using $\mathcal{H}^{\alpha}_j(k)=\mathbf{h}^{\alpha}_j\cdot{\pmb{\sigma}}$ and the Euler's formula of the Pauli matrices  \cite{PhysRevB.102.041119}
\begin{equation}
\begin{split}
e^{-i\mathbf{h}^\alpha_j\cdot\pmb{\sigma}T_j}&=\cos(E^\alpha_jT_j)-i\sin(E^\alpha_jT_j)\underline{\mathbf{h}}^\alpha_j\cdot\pmb{\sigma}
\end{split}
\end{equation}
with $E^\alpha_j=\sqrt{\mathbf{h}^\alpha_j\cdot\mathbf{h}^\alpha_j}$ and $\underline{\mathbf{h}}^\alpha_j=\mathbf{h}_j^\alpha/E_j^\alpha$, we have
\begin{eqnarray}
{U}_T&=&e^{-i\mathbf{h}^\alpha_2\cdot\pmb{\sigma} T_2}e^{-i\mathbf{h}^\alpha_1\cdot\pmb{\sigma} T_1}=\epsilon ^\alpha I_{2\times2}+i\mathbf{r}^\alpha\cdot\pmb{\sigma}\nonumber\\&&\equiv e^{-i\mathcal{H}^\alpha_\text{eff}T},\label{utt}\\
\epsilon^\alpha&=&\cos(E^\alpha_1T_1)\cos(E^\alpha_2T_2)-\underline{\mathbf{h}}^\alpha_1\cdot\underline{\mathbf{h}}^\alpha_2\sin(E^\alpha_2T_2)\nonumber\\&&\times\sin(E^\alpha_1T_1),~~~\label{epsl}\\
 \mathbf{r}^\alpha&=& \underline{\mathbf{h}}^\alpha_1\times\underline{\mathbf{h}}^\alpha_2\sin(T_1E^\alpha_1)\sin(T_2E^\alpha_2)-\underline{\mathbf{h}}^\alpha_2\cos(T_1E^\alpha_1)\nonumber\\ &&\times\sin(T_2E^\alpha_2)-\underline{\mathbf{h}}^\alpha_1\cos(T_2{E}^\alpha_2)\sin(T_1{E}^\alpha_1).\label{varepsilonds}
\end{eqnarray}
Using the Euler's formula again, we can infer $\mathcal{H}^\alpha_\text{eff}$ from Eq. \eqref{utt} as
\begin{eqnarray}
\mathcal{H}^\alpha_\text{eff}&=&-\arccos(\epsilon^\alpha)\mathbf{r}^\alpha\cdot\pmb{\sigma}/[\sin(\arccos(\epsilon^\alpha)T]\nonumber\\
&=&-\arccos(\epsilon^\alpha)\mathbf{r}^\alpha\cdot\pmb{\sigma}/[\sqrt{1-(\epsilon^\alpha)^2}T]\nonumber\\
&=&-\arccos(\epsilon^\alpha)\underline{\mathbf{r}}^\alpha\cdot\pmb{\sigma}/T,
\end{eqnarray}
where $(\epsilon^\alpha)^2+\mathbf{r}^\alpha\cdot \mathbf{r}^\alpha=1$ has been used. Thus we obtain $\mathbf{h}^\alpha_\text{eff}=-{\arccos \epsilon^\alpha\over T} \underline{\mathbf{r}}^\alpha$.

The quasienergy bands touch at $0$ and $\pi/T$, which occurs when $\epsilon^{\alpha}=+1$ and $-1$, respectively. Therefore, the phase transition is associated with the closing of the quasienergy bands, which occurs for ${ k}$ and driving parameters satisfying any one of the following conditions.
\begin{enumerate}
  \item $\sin(T_1E^\alpha_1)\sin(T_2E^\alpha_2)=0$. In this case, $\epsilon^\alpha=\cos(E^\alpha_1T_1)\cos(E^\alpha_2T_2)$. Then the bands of $\mathcal{H}^\alpha_\text{eff}$ close when
  \begin{equation}
T_jE^{\alpha}_j=n_j\pi,~n_j\in \mathbb{Z}. \label{gen}
  \end{equation}
  \item $\underline{\mathbf{h}}^{\alpha}_1\cdot\underline{\mathbf{h}}^{\alpha}_2=\pm1$. In this case, $\epsilon^\alpha=\cos(E^\alpha_1T_1\pm E^\alpha_2T_2)$. Then the bands of $\mathcal{H}^\alpha_\text{eff}$ close when
  \begin{equation}
  T_1{E}^{\alpha}_1\pm T_2{E}^{\alpha}_2=n\pi,~n\in\mathbb{Z},\label{hh1}
  \end{equation}
 at the quasienergy zero (or $\pi/T$) if $n$ is even (or odd).
\end{enumerate}

As the sufficient condition for the topological phase transition, Eqs. \eqref{gen} and \eqref{hh1} supply a guideline to manipulate the driving parameters for Floquet engineering to various topological phases at will.

\section{Winding number}
With the skin effect being removed, the bulk-boundary correspondence of the non-Hermitian system is retrieved. Then we can characterize its topological feature in the generalized Brillouin zone. Due to the mirror-rotation symmetry, the second-order topological insulator phases are sufficiently described by $\mathcal{H}(\tilde{k},\tilde{k})$ along the high-symmetry line $\tilde{k}_x=\tilde{k}_y\equiv\tilde{k}$ \cite{PhysRevLett.122.076801}. Diagonalizing $\mathcal{H}(\tilde{k},\tilde{k})$ into $\text{diag}[\mathcal{H}^+(\tilde{k}),\mathcal{H}^-(\tilde{k})]$ with $\mathcal{H}^\alpha(\tilde{k})={\bf h}^\alpha(\tilde{k})\cdot{\pmb\sigma}$ and
\begin{equation}
{\bf h}^\alpha(\tilde{k})=\sqrt{2}[v+\lambda \cos \tilde{k},\alpha (i\gamma+\lambda \sin \tilde{k}),0],\label{sblc}
\end{equation}
 we can describe its bulk topology by the mirror-graded winding number $\mathcal{W}=(\mathcal{W}_{+}-\mathcal{W}_{-})/2$. Here we define
$\mathcal{W}_\alpha=\frac{i}{2\pi}\int_{0}^{4\pi}\frac{\langle u^\text{L}_{\alpha}\lvert\partial_k\lvert u^\text{R}_{\alpha}\rangle}{\langle u^\text{L}_{\alpha}\lvert u^\text{R}_{\alpha}\rangle}dk$, where $\lvert u^\text{L,R}_{\alpha}\rangle$ are the right and left eigenstates of $\mathcal{H}^\alpha(\tilde{k})$ \cite{PhysRevLett.122.076801}.

Rewriting $\mathcal{H}^\alpha(\tilde{k})$ as
\begin{equation}
\mathcal{H}^\alpha(\tilde{k})=R^\alpha_{+}(\tilde{k})\sigma_{+}+R^\alpha_{-}(\tilde{k})\sigma_{-},
\end{equation}
where $\sigma_{\pm}=(\sigma_x\pm i\sigma_y)/2$, we can obtain their eigen values as
\begin{equation}
E^\alpha _{1}(\tilde{k})=-E^\alpha _{2}(\tilde{k})=\sqrt{R^\alpha_{+}(\tilde{k})R^\alpha_{-}(\tilde{k})}.
\end{equation}
The corresponding right and left eigenstates are
\begin{eqnarray}
\lvert u_{\alpha;1}^\text{R} \rangle&=&\frac{1}{\sqrt{2R^\alpha_{+}R^\alpha_{-}}}(R^\alpha_{+},\sqrt{R^\alpha_{+}R^\alpha_{-}})^{T},\\
\lvert u_{\alpha;2}^\text{R} \rangle&=&\frac{1}{\sqrt{2R^\alpha_{+}R^\alpha_{-}}}(R^\alpha_{+},-\sqrt{R^\alpha_{+}R^\alpha_{-}})^{T},\\
\langle u_{\alpha;1}^\text{L}\lvert&=&\frac{1}{\sqrt{2R^\alpha_{+}R^\alpha_{-}}}(R^\alpha_{-},\sqrt{R^\alpha_{+}R^\alpha_{-}}),\\
\langle u_{\alpha;2}^\text{L}\lvert&=&\frac{1}{\sqrt{2R^\alpha_{+}R^\alpha_{-}}}(R^\alpha_{-},-\sqrt{R^\alpha_{+}R^\alpha_{-}}).
\end{eqnarray}
Choosing the lower band $E^\alpha _{2}(\tilde{k})$, we can calculate the winding number
\begin{equation}
\begin{split}
\mathcal{W}_\alpha&=\frac{i}{2\pi}\int_{0}^{4\pi}\frac{\langle u_{\alpha,2}^\text{L}\lvert\partial_k\lvert u_{\alpha,2}^\text{R}\rangle}{\langle u_{\alpha,2}^\text{L}\lvert u_{\alpha,2}^\text{R}\rangle}dk=\frac{i}{4\pi}\int_{0}^{4\pi}d\ln p\\
 &=-\frac{1}{2\pi}\frac{[\arg R^\alpha_{+}-\arg R^\alpha_{-}]_{C_{\beta}}}{2},
\end{split}
\end{equation}
where $p=\sqrt{\frac{R_{+}}{R_{-}}}$ and $\arg R^\alpha_{\pm}$ are the phase changes of $R^\alpha_\pm$ as $\tilde{k}$ counterclockwisely goes along the GBZ $C_{\tilde{k}}$. From Eq. \eqref{sblc}, we have
\begin{eqnarray}
R^+_{+}&=&R^-_{-}=\sqrt{2}(v+\gamma+\lambda e^{- i\tilde{k}}),\\
R^+_{-}&=&R^-_{+}=\sqrt{2}(v-\gamma+\lambda e^{ i\tilde{k}}).
\end{eqnarray}
Then, we can readily obtain that, when $|\lambda|\geq|v^2-\gamma^2|$, the winding numbers $\mathcal{W}_{\pm}=\pm 1$. This result is consistent with the one under open boundary condition [see Figs. 1(b) and 1(c) in the main text].

%\newpage
\bibliography{references.bib}